\documentstyle[12pt]{article}
\catcode`@=11
\newcount\@tempcntc
\def\@citex[#1]#2{\if@filesw\immediate\write\@auxout{\string\citation{#2}}\fi
  \@tempcnta\z@\@tempcntb\m@ne\def\@citea{}\@cite{\@for\@citeb:=#2\do
    {\@ifundefined
       {b@\@citeb}{\@citeo\@tempcntb\m@ne\@citea\def\@citea{,}{\bf ?}\@warning
       {Citation `\@citeb' on page \thepage \space undefined}}%
    {\setbox\z@\hbox{\global\@tempcntc0\csname b@\@citeb\endcsname\relax}%
     \ifnum\@tempcntc=\z@ \@citeo\@tempcntb\m@ne
       \@citea\def\@citea{,}\hbox{\csname b@\@citeb\endcsname}%
     \else
      \advance\@tempcntb\@ne
      \ifnum\@tempcntb=\@tempcntc
      \else\advance\@tempcntb\m@ne\@citeo
      \@tempcnta\@tempcntc\@tempcntb\@tempcntc\fi\fi}}\@citeo}{#1}}
\def\@citeo{\ifnum\@tempcnta>\@tempcntb\else\@citea\def\@citea{,}%
  \ifnum\@tempcnta=\@tempcntb\the\@tempcnta\else
   {\advance\@tempcnta\@ne\ifnum\@tempcnta=\@tempcntb \else \def\@citea{--}\fi
    \advance\@tempcnta\m@ne\the\@tempcnta\@citea\the\@tempcntb}\fi\fi}
\catcode`@=12
\def\barr{\begin{array}}
\def\earr{\end{array}}
\def\beq{\begin{equation}}
\def\eeq{\end{equation}}
\def\bea{\begin{eqnarray}}
\def\eea{\end{eqnarray}}
\def\bmath{\begin{displaymath}}
\def\emath{\end{displaymath}}
\def\bq{\begin{quote}}
\def\eq{\end{quote}}
\def\slash#1{\setbox0=\hbox{$#1$}#1\hskip-\wd0\hbox to\wd0{\hss\sl/\/\hss}}
\begin{document}

\begin{flushright}
MAD/PH/787 \\
MZ-TH/93-22 \\
August 1993
\end{flushright}

\begin{center}
{\bf{\Large {\em CP} Violation Induced by Heavy Majorana}}\\[0.3cm]
{\bf{\Large  Neutrinos}} \\[0.3cm]
{\bf{\Large in the Decays of Higgs Scalars into}} \\[0.3cm]
{\bf{\Large Top-Quark, {\em W}- and {\em Z}-Boson Pairs }}\\[2.cm]
{\large A.~Ilakovac}$^{a}$,
{\large B.A.~Kniehl}$^{b}$
\footnote[1]{On leave from II. Institut f\"ur Theoretische Physik,
Universit\"at Hamburg, 22761 Hamburg, Germany.}
{\large and A.~Pilaftsis}$^{a}$
\footnote[2]{Address after 1 October~1993: Rutherford Appleton Laboratory,
Chilton, Didcot, Oxon, England.}
\footnote[3]{E-mail address: pilaftsis@vipmza.physik.uni-mainz.de}\\[0.4cm]
$^{a}$ Institut f\"ur Physik, Johannes-Gutenberg Universit\"at,
55099 Mainz, Germany\\[0.3cm]
$^{b}$ Deptartment of Physics, University of Wisconsin, Madison, WI 53706,
USA
\end{center}
\bigskip
\bigskip
\bigskip
\bigskip
\centerline {\bf ABSTRACT}

We analyze the possibility of $CP$ violation induced by heavy
Majorana neutrinos in the decays of the Higgs particle into top-quark, $W$- and
$Z$-boson pairs. In the framework of various ``see-saw" models with interfamily
mixings, we find that Majorana neutrinos may give rise to sizable $CP$-odd
observables at the one-loop electroweak order. Numerical estimates of these
$CP$-violating effects that may be detected in high-energy colliders
are presented.

\newpage

The possible existence of heavy Majorana neutrinos plays an important r\^ole in
addressing a number of outstanding questions in cosmology and
astrophysics~\cite{ASTR} like the smallness in mass of the known neutrinos,
the solar neutrino deficit~\cite{MSW}, the baryon asymmetry in the
universe~\cite{VAY}, etc.
Apart from the possibility of producing such heavy neutral leptons at high
energy $ee$, $ep$, or $pp$ colliders~\cite{PROD}, their presence
may be manifestated by detecting lepton-flavour violating
decays of the $Z$ and the Higgs ($H^0$) particle~\cite{AP,KPS,JB,BP}
or through their influence~\cite{KK} on the electroweak oblique parameters,
$S$, $T$, $U$, or $\varepsilon_1$, $\varepsilon_2$,
$\varepsilon_3$~\cite{PT,DKS,HKS}.

Another interesting aspect, which will be investigated in this letter,
is that quantum effects mediated by heavy Majorana neutrinos can induce
sizable $CP$-odd signals~\cite{APmaj} in the production of top-quark ($t$),
$W$- and $Z$-boson pairs originating from the decay of the Standard Model
($SM$) Higgs boson.
Specifically, we shall investigate whether $CP$ asymmetries defined
as~\cite{Peskin,CH,PN}
\beq
A_{CP}^{(t)}\ =\ \frac{\Gamma(H^0 \to t_L \bar{t}_L)\ -\
\Gamma(H^0 \to t_R \bar{t}_R) }{\Gamma(H^0 \to t \bar{t}\,) }\
\eeq 
and~\cite{CH}
\beq
A_{CP}^{(W)}\ =\ \frac{\Gamma(H^0 \to W^+_{(+1)}W^-_{(+1)})\ -\
\Gamma(H^0 \to W^+_{(-1)}W^-_{(-1)})}
{\Gamma(H^0 \to W^+_{(+1)}W^-_{(+1)})\ +\ \Gamma(H^0 \to W^+_{(-1)}W^-_{(-1)})}
\eeq 
may be observable at $LHC$ or $SSC$ energies. The subscripts $L,R$,
and $(\pm 1)$
in Eqs.~(1) and~(2) denote the two helicity states of the top quark and
the helicities of the transverse degrees of the $W$ boson, respectively.
The states $t_L$ and $W^-_{(+1)}$ are connected with the states $\bar{t}_R$
and $W^+_{(-1)}$ by $CP$ conjugation.
Therefore, $A_{CP}^{(t)}$ and $A_{CP}^{(W)}$ represent
genuine $CP$-violating parameters that can be determined experimentally.
By analogy to Eq.~(2), a $CP$ asymmetry, $A_{CP}^{(Z)}$, for the decay
channel $H^0 \to ZZ$ can be constructed. It should
be stressed that the $CP$ asymmetries $A_{CP}^{(t)}$, $A_{CP}^{(W)}$,
and $A_{CP}^{(Z)}$ resulting from the Feynman graphs shown in Fig.~1 cannot
be induced if the heavy neutrinos are of standard Dirac type.
To be more specific, at one loop in the SM, the amplitude of $H^0\to W^+W^-$
contains a term proportional to
$\epsilon_{\mu\nu\rho\sigma}\varepsilon^\mu(k_+)\varepsilon^\nu(k_-)k_+^\rho
k_-^\sigma$, where $k_\pm$ and $\varepsilon(k_\pm)$ denote the
four-momenta and polarization four-vectors of the $W^\pm$ bosons,
respectively.
However, the coefficient of this term vanishes for $k_+^2=k_-^2$ and,
in particular, for on-shell $W$ bosons~\cite{hww}.

The minimal class of models that predict heavy Majorana neutrinos can be
obtained by simply adding $n_G$ right-handed neutrinos, $\nu^0_{R_i}$,
to the field content of the $SM$~\cite{VS}, where $n_G$ denotes the number
of generations. After spontaneous break-down
of the $SU(2)\otimes U(1)$ gauge symmetry, the Yukawa sector containing the
neutrino masses reads
\beq
- {\cal L}_Y^\nu\ =\ \frac{1}{2} (\bar{\nu}^0_L,\ \bar{\nu}^{0C}_R)
M^\nu \left( \barr{c} \nu^{0C}_L \\ \nu^0_R \earr \right)\ \quad +
\quad H.c.\ ,
\eeq 
where the neutrino mass matrix,
\beq
M^\nu\ \ =\ \ \left( \barr{cc} 0 & m_D \\
                              m_D^T & m_M \earr \right) ,
\eeq 
takes the known ``see-saw" form~\cite{YAN}. In Eq.~(4), $m_D$ and $m_M$ are
$n_G\times n_G$-dimensional matrices that
cannot, in general, be diagonalized simultaneously. Note also that
the $B-L$-breaking mass terms, $m_{M_{ij}}$, can be assumed to be bare masses
in the Lagrangian without violating the gauge symmetry of the $SM$.
The symmetric
matrix $M^\nu$ can be brought into diagonal form by a $2n_G\times 2n_G$
unitary matrix $U^\nu$ (i.e., $\hat{M}^\nu=U^{\nu T} M^\nu U^\nu$). This yields
$2n_G$ physical Majorana neutrinos, $n_i$, which are related with the
weak eigenstates through the unitary transformations
\beq
\left( \barr{c} \nu_L^0 \\ \nu^{0C}_R \earr \right)_i \ =\
U^{\nu\ast}_{ij}\ n_{L_j}\ , \qquad
\left( \barr{c} \nu_L^{0C} \\ \nu^0_R \earr \right)_i \ =\
U^\nu_{ij}\ n_{R_j} \ .
\eeq 
The first $n_G$ Majorana neutrinos, $\nu_i$, are identified with the ordinary
neutrinos, $\nu_e$, $\nu_\mu$, etc.\
(i.e., $\nu_i=n_i$ for $i=1,\dots,n_G$),
while the remaining $n_G$ mass eigenstates represent heavy Majorana neutrinos
that are predicted by the model (i.e., $N_i=n_{i+n_G}$).
It is now straightforward
to obtain for our purposes the relevant interactions of the Majorana
neutrinos with the $W$-, $Z$-, and Higgs bosons in this minimal model.
Adopting the notation of~\cite{ZPC}, these interactions are given
in a renormalizable form by
\bea
{\cal L}_{int}^W &=& -\ \frac{g_W}{2\sqrt{2}} W^{-\mu}\
{\bar{l}}_i \ B_{l_ij} {\gamma}_{\mu} (1-{\gamma}_5) \ n_j \quad + \quad H.c.\
,
\\[0.3cm]
{\cal L}_{int}^Z &=& -\ \frac{g_W}{4\cos\theta_W}  Z^\mu\
\bar{n}_i \gamma_\mu \Big[ i\mbox{Im}C_{ij}\ -\ \gamma_5\mbox{Re}C_{ij}
\Big] n_j\ ,\\[0.3cm]
{\cal L}^H_{int} &=& -\ \frac{g_W}{4M_W}\
H^0\ \bar{n}_i \Bigg[ (m_{n_i}+m_{n_j})\mbox{Re}C_{ij}
+\ i\gamma_5 (m_{n_j}-m_{n_i})\mbox{Im}C_{ij} \Bigg] n_j\ ,
\eea 
where
\bea
C_{ij}\ &=&\ \sum\limits_{k=1}^{n_G}\ U^\nu_{ki}U^{\nu\ast}_{kj}\ ,\\
B_{l_ij}\ &=& \sum\limits_{k=1}^{n_G} V^l_{l_ik} U^{\nu\ast}_{kj}\ .
\eea 
The matrices $B$ and $C$ obey a great number of useful identities that
will help us to quantify our $CP$ asymmetries. These identities
read~\cite{AP,KPS}
\bea
\sum\limits_{i=1}^{2n_G} B_{l_1i}B_{l_2i}^{\ast} & = & \ {\delta}_{l_1l_2},\\
\sum\limits_{k=1}^{2n_G} C_{ik}C^\ast_{jk} & = & \ C_{ij}, \\
\sum\limits_{i=1}^{2n_G} B_{li}C_{ij} & = & \ B_{lj}, \\
\sum\limits_{k=1}^{2n_G} m_{n_i} C_{ik}C_{jk} & = & \ 0, \\
\sum\limits_{k=1}^{n_G} B_{l_ki}^{\ast}B_{l_kj} & = & \ C_{ij}.
\eea
In general, the mixings between light-heavy states, i.e., $B_{lN}$ or
$C_{\nu N}$, can be constrained by a global analysis of low-energy
experiments and $LEP$ data~\cite{LL}.

It is worth mentioning that the Lagrangians of Eqs.~(6)--(8) violate the $CP$
symmetry of the model. Analytically, Eq.~(6) introduces $CP$ violation
by the
known complex Cabbibo-Kobayashi-Maskawa-type matrix, $B_{l_in_j}$.
The neutral-current interactions described by  Eqs.~(7) and~(8) violate
also the $CP$ symmetry of the model, since the neutral particles $H^0$ and $Z$
couple simultaneously to $CP$-even ($\bar{n}_in_j$ or $\bar{n}_i\gamma_\mu
n_j$) and $CP$-odd
($\bar{n}_i\gamma_5n_j$ or $\bar{n}_i\gamma_\mu\gamma_5 n_j$) operators with
two different Majorana neutrinos (i.e., $n_i\neq n_j$)~\cite{Marek}.
As a consequence of the latter, non-zero $CP$-odd parts are generated
radiatively in the $H^0t\bar{t}$, $H^0W^+W^-$, and $H^0ZZ$ couplings,
and thus $CP$ violation is induced in the corresponding decays of the
Higgs particle (see also Fig.~1).

After setting the stage, we are now in a position to calculate the $CP$
asymmetry related to the $H^0\to t\bar{t}$ decay.
Actually, we are looking for $CP$-odd correlations of the
type $\langle(\vec{s}_t - \vec{s}_{\bar{t}} )\cdot\vec{k}_t\rangle$ as given in
Eq.~(1). This kind of $CP$-odd observables, being odd under $CPT$, should
combine with the $CPT$-odd
absorptive parts in the totally integrated decay rates.
The important ingredient for a non-zero $A_{CP}^{(t)}$ are Eqs.~(7) and (8),
which violate the $CP$ symmetry of the model, as pointed out earlier.
Thus, vacuum-polarization transitions between the $CP$-even Higgs and
the $CP$-odd $Z$ boson, which are forbidden in the $SM$, can now
give rise to a pseudoscalar part in the $H^0t\bar{t}$ coupling (see
also Fig.~1(a)). In this way, one obtains
\beq
A_{CP}^{(t)}\ =\ \frac{\alpha_W}{4}\ \mbox{Im}C^2_{ij}
\sqrt{\lambda_i\lambda_j}
\ \frac{\lambda_i-\lambda_j}{\lambda_H}\
\frac{\lambda^{1/2}(\lambda_H,\lambda_i,\lambda_j)}
{\lambda^{1/2}(\lambda_H,\lambda_t,\lambda_t)}\ ,
\eeq 
where
\bea
\lambda_i \ = \ \frac{m_{n_i}^2}{M^2_W}, \qquad \lambda_H\ =\
\frac{M^2_H}{M^2_W}, \qquad \lambda_t\ =\ \frac{m^2_t}{M^2_W}, \nonumber\\
\lambda(x,y,z)\ =\ (x-y-z)^2-4yz.
\eea 
{}From Eq.~(16) we see that two non-degenerate Majorana neutrinos with
appreciable masses are at least
required in order to get $A_{CP}^{(t)}\neq 0$. In general, the number of
$CP$-odd phases that exist theoretically in the $SM$ with right-handed
neutrinos is~\cite{KPS}
\beq
{\cal N}_{CP}\ =\ N_L(N_R\ -\ 1),
\eeq 
where $N_L$ and $N_R$ are the numbers of left-handed and right-handed
neutrinos,
respectively. Assuming now three generations, one has
Im$C^2_{N_1N_2} \leq 10^{-2}$~\cite{LL}. However, the situation changes if
one introduces an additional left-handed neutrino field, $\nu^0_{L_4}$, in the
Lagrangian of the $SM$~\cite{PH}. If a non-trivial mixing between the two
right-handed neutrinos is assumed, $M^\nu$ of Eq.~(4) takes the form
\beq
M^\nu\ =\ \left( \barr{ccc}
0 & a & b\\
a & A & 0 \\
b & 0 & B  \earr \right).
\eeq 
Obviously, the mass eigenvalues of $M^\nu$ should be in excess of $M_Z/2$
in order for te respective particles to escape detection at $LEP$
experiments. Employing Eq.~(14), one
can derive the helpful relations
\bea
\mbox{Im}C^2_{\nu_4N_1}\ &=&\ \sin\delta_{CP}\ |C_{\nu_4N_1}|^2,\\
\mbox{Im}C^2_{\nu_4N_1}\ &=&\ -\frac{m_{N_1}}{m_{N_2}}\sin\delta_{CP}\
|C_{\nu_4N_1}|^2, \\
\mbox{Im}C^2_{N_1N_2}\ &=&\ \frac{m_{\nu_4}}{m_{N_2}}\sin\delta_{CP}\
|C_{\nu_4N_1}|^2.
\eea 
In these models, the mixing $|C_{\nu_4N_1}|^2$ can be of order one. In
Table~1, we present the numerical results for $A_{CP}^{(t)}$ within the two
models mentioned above. In particular, we see that $CP$ asymmetries
$A_{CP}^{(t)} \simeq 4.\ 10^{-2}$ are conceivable in such four-generation
scenarios with Majorana neutrinos~\cite{PH,IKP}.
We emphasize again the fact that high-mass
Dirac neutrinos cannot produce a non-zero $CP$ asymmetry.

Similarly, heavy Majorana neutrinos can generate a $CP$-odd part
in the $H^0W^+W^-$ coupling through the triangle graph shown in Fig.~1(b).
Since our model is free of anomalies~\cite{NP}, one can uniquely search
for $CP$-violating correlations of the form
\beq
\epsilon_{\mu\nu\rho\sigma}\ \varepsilon_{(+1)}^\mu(\hat{z})
\varepsilon_{(+1)}^\nu(-\hat{z}) k_+^\rho k_-^\sigma\ =
\ M_H \vec{k}_+\ (\vec{\varepsilon}_{(+1)}(\hat{z}) \ \times
\ \vec{\varepsilon}_{(+1)}(-\hat{z})),
\eeq 
In Eq.~(23), we have assumed that the polarization vectors,
$\varepsilon_{(+1)}^\mu(\hat{z})$ and $\varepsilon_{(+1)}^\nu(-\hat{z})$,
describe two transverse $W$ bosons with helicity $+1$. The presence of
$CP$-odd terms in the transition amplitude of the decay $H^0\to W^+W^-$
leads to the $CP$ asymmetry
\beq
A_{CP}^{(W)}\  =\ \frac{\alpha_W}{4} \Big[ \mbox{Im}(B_{l_ki}
B_{l_kj}^\ast C_{ij}^\ast)\sqrt{\lambda_i\lambda_j}\ (
F_+\ +\ F_- )\ +\ \mbox{Im}(B_{l_ki}B_{l_kj}^\ast C_{ij})
(\lambda_i F_+\ +\ \lambda_j F_-) \ \Big],
\eeq 
where $\alpha_W=(g_W^2/4\pi)$.
The functions $F_\pm$ in Eq.~(24) receive absorptive contributions from
three different kinematic configurations of the intermediate states that
can become on-shell. A straightforward computation of all absorptive
parts shown in Fig.~1(b) gives
\bea
F_\pm(\lambda_i,\lambda_j, \lambda_{l_k}) &=&
\theta(M_H-m_{n_i}-m_{n_j})\Bigg[ \ \pm\ \frac{\lambda^{1/2}
(\lambda_H,\lambda_i,\lambda_j)}{2\lambda^{1/2}(\lambda_H,\lambda_W,\lambda_W)}
\ +\ G_\pm(\lambda_i,\lambda_j,\lambda_{l_k})\nonumber\\
&& \times\ln\left( \frac{t^+(\lambda_i,\lambda_j)-\lambda_{l_k}}{
t^-(\lambda_i,\lambda_j)-\lambda_{l_k}}\right)\ \Bigg]\
+\ \theta(M_W-m_{n_i}-m_{l_k}) \Bigg[\Bigg(\ - \frac{\lambda_H-4\lambda_W}
{\lambda_W}\nonumber\\
&& \mp\ \frac{\lambda_H}{\lambda_W}\ \Bigg)\frac{\lambda^{1/2}(
\lambda_W,\lambda_i,\lambda_{l_k})}{4\lambda^{1/2}(\lambda_H,
\lambda_W,\lambda_W)}\ +\ G_\pm(\lambda_i,\lambda_j,\lambda_{l_k})
\ln\left( \frac{\bar{t}^+(\lambda_i,\lambda_{l_k})-\lambda_j}
{\bar{t}^-(\lambda_i,\lambda_{l_k})-\lambda_j}\right)\ \Bigg]\nonumber\\
&&+\ \theta(M_W-m_{n_j}-m_{l_k})\Bigg[\left(\ \frac{\lambda_H-4\lambda_W}
{\lambda_W}\ \mp\ \frac{\lambda_H}{\lambda_W}\ \right)\frac{\lambda^{1/2}(
\lambda_W,\lambda_j,\lambda_{l_k})}{4\lambda^{1/2}(\lambda_H,
\lambda_W,\lambda_W)}\nonumber\\
&&+\ G_\pm(\lambda_i,\lambda_j,\lambda_{l_k})
\ln\left( \frac{\bar{t}^+(\lambda_j,\lambda_{l_k})-\lambda_i}
{\bar{t}^-(\lambda_j,\lambda_{l_k})-\lambda_i}\right)\ \Bigg],
\eea 
where
\bea
\lambda_{l_k} &=& \frac{m_{l_k}^2}{M^2_W}, \qquad \lambda_W\ =\ 1,
\qquad \lambda_Z\ =\ \frac{M^2_Z}{M^2_W},\\
t^\pm(x,y) &=&
-\frac{1}{2}\Big[ \ \lambda_H-x-y-2\lambda_W
\ \mp\ \lambda^{1/2}(\lambda_H,x,y)
\lambda^{1/2}(\lambda_H,\lambda_W,\lambda_W)\  \Big],\\
\bar{t}^\pm(x,y)&=& \lambda_H+x - \frac{1}{2}\lambda_H(\lambda_W
+x-y)\ \pm \frac{1}{2}\lambda^{1/2}(\lambda_H,x,y)\lambda^{1/2}(\lambda_H,
\lambda_W,\lambda_W),\\
G_\pm(x,y,z)&=&\frac{x-y}{4\lambda_H}\ \pm\ \frac{2(\lambda_W+z)-
x-y}{4(\lambda_H-4\lambda_W)}.
\eea 
The dominant contribution to $A_{CP}^{(W)}$ comes from the $n_in_j$ on-shell
states, while absorptive parts arising from $n_il_k$ or $n_jl_k$ states are
vanishingly small. In Table~2, we exhibit numerical estimates for
$A_{CP}^{(W)}$ within the two representative ``see-saw" scenarios discussed
above.
In order to measure $A_{CP}^{(W)}$,
one has to be able to discriminate the events
$W^+_{(+1)}W^-_{(+1)}$ and $W^+_{(-1)}W^-_{(-1)}$ from the total number of
$W$ bosons produced by Higgs-boson decays. An estimate can be obtained from
the following ratio:
\bea
R^{(W)}\ &=&\ \frac{\Gamma(H^0 \to W^+_{(+1)}W^-_{(+1)})\ +\
\Gamma(H^0 \to W^+_{(-1)}W^-_{(-1)})}{\Gamma (H^0\to W^+W^-)}\nonumber\\
&=& \frac{8M_W^4}{M^4_H}\ \left( \  1 \ -\ 4\frac{M^2_W}{M^2_H}\
+\ 12\frac{M^4_W}{M^4_H}\ \right)^{-1}.
\eea 
Since one expects to be able to analyze $10^5$--$10^6$ Higgs decays per year
at $SSC$ or $LHC$ for a wide range
of Higgs-boson masses, i.e., $M_H=300$--$800$~GeV,
it should, in principle, be possible to see
$R^{(W)}$ values of order $10^{-2}$--$10^{-3}$
and $CP$ asymmetries of $A_{CP}^{(W)} \simeq 10\%$
in the $SM$ with three (four) generations.

Similarly, heavy Majorana neutrinos induce, through the triangle graph
of Fig.~1(c), a non-zero $CP$ asymmetry in the decay $H^0\to ZZ$,
which is found to be
\bea
A_{CP}^{(Z)} \ &=&\ \frac{\alpha_W}{4}\ \Bigg[
\mbox{Im}(C_{ij}C_{jk}C_{ki})\ ( \lambda_iF_1\ +\ \lambda_jF_2 )
\ +\ \mbox{Im}(C_{ij}^\ast C_{jk}C_{ki})\sqrt{\lambda_i\lambda_j}
(F_1\ +\ F_2)\nonumber\\
&&-\ \Big(\ \mbox{Im}(C_{ij}^\ast C_{jk}^\ast C_{ki})\sqrt{\lambda_i\lambda_k}
\ +\ \mbox{Im}(C_{ij}C_{jk}^\ast C_{ki})\sqrt{\lambda_j\lambda_k}\ \Big)
(R\ +\ F_1\ -\ F_2) \Bigg],
\eea 
where
\bea
R\ & =&\ \theta (M_H-m_{n_i}-m_{n_j})\ \frac{1}{2}\
\ln\left( \frac{t^+(\lambda_i,\lambda_j)-\lambda_k}
{t^-(\lambda_i,\lambda_j)-\lambda_k} \right)\nonumber\\
&&+\ \theta (M_W-m_{n_i}-m_{n_k})\ \frac{1}{2}\
\ln\left( \frac{\bar{t}^+(\lambda_i,\lambda_k)-\lambda_j}
{\bar{t}^-(\lambda_i,\lambda_k)-\lambda_j} \right)\nonumber\\
&&+\ \theta (M_W-m_{n_j}-m_{n_k})\ \frac{1}{2}\
\ln\left( \frac{\bar{t}^+(\lambda_j,\lambda_k)-\lambda_i}
{\bar{t}^-(\lambda_j,\lambda_k)-\lambda_i} \right).
\eea 
The functions $F_1$, $F_2$, $t^\pm$, and $\bar{t}^\pm$ are obtained
from $F_+$, $F_-$, Eqs.~(27) and (28), respectively, by making the
obvious replacements, $\lambda_W\to \lambda_Z$ and $\lambda_{l_k} \to
\lambda_k$, in the relevant formulae. From Table 3 we see
that $A_{CP}^{(Z)}$ can be of the order of $10\%$, which should be sizable
enough to observe such a $CP$-violating effect at $pp$ supercolliders.

In conclusion, we have shown that heavy Majorana neutrinos
are of potential interest in accounting for possible $CP$-violating
phenomena in the decays of Higgs bosons into $t\bar{t}$, $W^+W^-$, and $ZZ$
pairs.
The minimal extension of the $SM$ by right-handed neutrinos
can naturally account for sizable
$CP$-violating effects of the order of $10\%$ at high-energy colliders
and thus provides an attractive alternative to complicated
multi-Higgs-boson scenarios tuned to yield large $CP$-odd
effects~\cite{Peskin,CH,PN,CPhig}.
We must remark that the $CP$ asymmetries $A_{CP}^{(t)}$,
$A_{CP}^{(W)}$, and $A_{CP}^{(Z)}$ may not be directly accessible
by experiment.
However, the $CP$-violating signals originating from such Higgs-boson decays
will be transcribed to the decay products of the top-quark, $W$-, and
$Z$-boson pairs.
More realistic $CP$-odd projectors can be constructed, for instance,
by considering angular-momentum distributions or energy asymmetries of the
produced charged leptons and jets~\cite{Peskin,CH,BAK}.
\\[2cm]
{\bf Acknowledgements.}  We wish to thank X.~Zhang and A.~Mukhamedzhanov
for discussions.
$BK$ is grateful to the Phenomenology Department at the University of
Wisconsin-Madison for the warm hospitality.
The work of $BK$ has been supported in part
by the U.S. Department of Energy under Contract No.\ DE-AC02-76ER00881,
in part by the Texas National Laboratory Research Commission
under Grant No.\ RGFY93-221, and in part by the
University of Wisconsin Research Committee with funds granted by the
Wisconsin Alumni Research Foundation.
The work of $AP$ has been supported by
a grant from the Postdoctoral Graduate College of Mainz.

\newpage

\centerline{\bf\Large Figure and Table Captions }
\vspace{1cm}
\newcounter{fig}
\begin{list}{\bf\rm Fig. \arabic{fig}: }{\usecounter{fig}
\labelwidth1.6cm \leftmargin2.5cm \labelsep0.4cm \itemsep0ex plus0.2ex }

\item Feynman graphs giving rise to a $CP$-odd part in the
$H^0t\bar{t}$, $H^0W^+W^-$, and $H^0ZZ$ couplings.

\end{list}
\newcounter{tab}
\begin{list}{\bf\rm Tab. \arabic{tab}: }{\usecounter{tab}
\labelwidth1.6cm \leftmargin2.5cm \labelsep0.4cm \itemsep0ex plus0.2ex }

\item Numerical results of $A^{(t)}_{CP}/\mbox{Im}C^2_{N_1N_2}$
($A^{(t)}_{CP}/\mbox{Im}C^2_{\nu_4N_1}$) in the context of interfamily
``see-saw" models with three (four) generations.
We use $m_{N_1}\ (m_{\nu_4})=100$~GeV
for the one heavy neutrino and $m_t=150$~GeV.

\item Numerical values of $A^{(W)}_{CP}/\mbox{Im}C^2_{N_1N_2}$ and
$A^{(W)}_{CP}/\mbox{Im}C^2_{\nu_4N_1}$ for ``see-saw" models with three and
four generations, respectively.
We use $m_l=0$ for the charged leptons of the first three generations
and $m_E=100$~GeV~\cite{PDG} for the charged lepton of the
fourth generation.

\item Numerical results of $A^{(Z)}_{CP}/\mbox{Im}C^2_{N_1N_2}$ and $
A^{(Z)}_{CP}/\mbox{Im}C^2_{\nu_4N_1}$ for ``see-saw" models with three and
four generations, respectively.

\end{list}

\newpage

\centerline{\bf\Large Table 1}
\vspace{1.5cm}
\begin{tabular*}{11.22cm}{|r||c|c|c|}
\hline
 & & &  \\
$m_{N_2}$ & $M_H=400$~GeV &  $M_H=600$~GeV & $M_H=800$~GeV \\
$[$GeV$]$ & & &\\
\hline\hline
&&& \\
150 & -3.78~$10^{-3}$ & -1.50~$10^{-3}$ & -8.25~$10^{-4}$ \\
200 & -1.00~$10^{-2}$ & -4.52~$10^{-3}$ & -2.56~$10^{-3}$ \\
250 & -1.53~$10^{-2}$ & -9.12~$10^{-3}$ & -5.38~$10^{-3}$ \\
300 &     $-$         & -1.49~$10^{-2}$ & -9.34~$10^{-3}$ \\
400 &     $-$         & -2.54~$10^{-2}$ & -2.01~$10^{-2}$ \\
450 &     $-$         & -2.48~$10^{-2}$ & -2.63~$10^{-2}$ \\
475 &     $-$         &     $-$         & -2.92~$10^{-2}$ \\
500 &     $-$         &     $-$         & -3.19~$10^{-2}$ \\
600 &     $-$         &     $-$         & -3.68~$10^{-2}$ \\
650 &     $-$         &     $-$         & -3.14~$10^{-2}$ \\

&&&\\
\hline
\end{tabular*}

\newpage

\hoffset=+1cm
\centerline{\bf\Large Table 2}
\vspace{1.5cm}
\begin{tabular*}{16.41cm}{|r||rr|rr|rr|}
\hline
 & & & & & & \\
$m_{N_2}$ & $M_H$&$=\ \ 400$~GeV &  $M_H$&$=\ \ 600$~GeV & $M_H$&$=\ \
800$~GeV  \\
$[$GeV$]$& $m_l\quad $& $m_E\quad $& $m_l\quad$& $m_E\quad$& $m_l\quad$&
$m_E\quad$\\
& 0~GeV & 100~GeV & 0~GeV & 100~GeV & 0~GeV & 100~GeV \\
\hline\hline
&&&&&& \\
150 & -8.61~$10^{-3}$ & -3.98~$10^{-3}$ & -5.33~$10^{-3}$
    & -2.48~$10^{-3}$ & -3.52~$10^{-3}$ & -1.65~$10^{-3}$ \\
200 & -2.11~$10^{-2}$ & -1.13~$10^{-2}$ & -1.49~$10^{-2}$
    & -7.76~$10^{-3}$ & -1.02~$10^{-2}$ & -5.23~$10^{-3}$ \\
250 & -3.13~$10^{-2}$ & -1.94~$10^{-2}$ & -2.87~$10^{-2}$
    & -1.64~$10^{-2}$ & -2.04~$10^{-2}$ & -1.13~$10^{-2}$ \\
300 &  0              &  0              & -4.58~$10^{-2}$
    & -2.83~$10^{-2}$ & -3.44~$10^{-2}$ & -2.02~$10^{-2}$ \\
400 &  0              &  0              & -7.90~$10^{-2}$
    & -5.65~$10^{-2}$ & -7.23~$10^{-2}$ & -4.76~$10^{-2}$ \\
450 &  0              &  0              & -8.02~$10^{-2}$
    & -6.14~$10^{-2}$ & -9.46~$10^{-2}$ & -6.55~$10^{-2}$ \\
475 &  0              &  0              & -6.71~$10^{-2}$
    & -5.29~$10^{-2}$ & -0.106          & -7.51~$10^{-2}$ \\
500 &  0              &  0              &  0
    &  0              & -0.117          & -8.48~$10^{-2}$ \\
600 &  0              &  0              &  0
    &  0              & -0.147          & -0.115           \\
650 &  0              &  0              &  0
    &  0              & -0.132          & -0.109           \\
&&&&&&\\
\hline
\end{tabular*}
\hoffset=-1cm

\newpage
\hoffset=-1cm

\centerline{\bf\Large Table 3}
\vspace{1.5cm}
\begin{tabular*}{16.805cm}{|r||cc|cc|cc|}
\hline
 & & & &&& \\
$m_{N_3}$ & $M_H$&$=\quad 400$~GeV &  $M_H$&$=\quad 600$~GeV & $M_H$&$=\quad
800$~GeV \\
$[$GeV$]$& $m_{N_1}(m_{\nu_4})$&$=\quad 150$~GeV& $m_{N_1}(m_{\nu_4})$&$=\quad
250$~GeV & $m_{N_1}(m_{\nu_4})$&$=\quad 350$~GeV  \\
& $m_{N_2}$&$=\quad 250$~GeV & $m_{N_2}$&$=\quad 350$~GeV &$m_{N_2}$&$=\quad
450$~GeV  \\
 &$3Gens$&$4Gens$&$3Gens$&$4Gens$&$3Gens$&$4Gens$ \\
\hline\hline
&&&&&& \\
300 &-1.60~$10^{-2}$ &-1.41~$10^{-2}$ &
    &                &                &                \\
400 &-3.70~$10^{-2}$ &-3.23~$10^{-2}$ &-2.93~$10^{-2}$
    &-2.29~$10^{-2}$ &                &                \\
450 &-4.40~$10^{-2}$ &-3.83~$10^{-2}$ &-5.32~$10^{-2}$
    &-4.15~$10^{-2}$ &                &                \\
500 &-4.95~$10^{-2}$ &-4.29~$10^{-2}$ &-7.30~$10^{-2}$
    &-5.67~$10^{-2}$ &-4.28~$10^{-2}$ &-3.00~$10^{-2}$  \\
600 &-5.74~$10^{-2}$ &-4.94~$10^{-2}$ &-0.103
    &-7.97~$10^{-2}$ &-0.111          &-7.78~$10^{-2}$  \\
650 &-6.03~$10^{-2}$ &-5.17~$10^{-2}$ &-0.115
    &-8.84~$10^{-2}$ &-0.138          &-9.68~$10^{-2}$  \\
1000&-7.09~$10^{-2}$ &-6.02~$10^{-2}$ &-0.161
    &-0.122          &-0.255          &-0.178           \\
&&&&&&\\
\hline
\end{tabular*}


\begin{thebibliography}{99}
\bibitem{IKP} In fact, the model constructed here differs crucially from
that of~\cite{PH}. One should require a strong mixing between $\nu^0_{L_4}$
and two right-handed neutrinos, $\nu^0_{R_1}$ and $\nu^0_{R_2}$, in order to
describe $CP$ violation. As a consequence of the absence of interfamily
mixings, the Hill-Paschos model corresponds effectively to the case $N_R=1$,
which implies ${\cal N}_{CP}=0$ on account of Eq.~(18).
\end{thebibliography}
\end{document}